\documentclass[notitlepage,a4paper,aps,prd,onecolumn,superscriptaddress,nofootinbib,groupedaddress,longbibliography]{revtex4-1}
\usepackage{amsmath}
\usepackage{amsfonts}
\usepackage{amssymb}
\usepackage[utf8]{inputenc}
\usepackage[T1]{fontenc}
\usepackage[colorlinks=true]{hyperref}
\usepackage[dvipsnames]{xcolor}
\usepackage{graphicx}
\usepackage[normalem]{ulem}

\newcommand{\D}[0]{\mathcal{D}}
\newcommand{\dd}[0]{\mathrm{d}}

\begin{document}

\title{Hamiltonian and primary constraints of new general relativity}

\author{Daniel Blixt}
\email{blixt@ut.ee}
\affiliation{Laboratory of Theoretical Physics, Institute of Physics, University of Tartu, W. Ostwaldi 1, 50411 Tartu, Estonia}

\author{Manuel Hohmann}
\email{manuel.hohmann@ut.ee}
\affiliation{Laboratory of Theoretical Physics, Institute of Physics, University of Tartu, W. Ostwaldi 1, 50411 Tartu, Estonia}

\author{Christian Pfeifer}
\email{christian.pfeifer@ut.ee}
\affiliation{Laboratory of Theoretical Physics, Institute of Physics, University of Tartu, W. Ostwaldi 1, 50411 Tartu, Estonia}

\begin{abstract}
We derive the kinematic Hamiltonian for the so-called ``new general relativity'' class of teleparallel gravity theories, which is the most general class of theories whose Lagrangian is quadratic in the torsion tensor and does not contain parity violating terms. Our approach makes use of an explicit expression for the flat, in general, nonvanishing spin connection, which avoids the use of Lagrange multipliers, while keeping the theory invariant under local Lorentz transformations. We clarify the relation between the dynamics of the spin connection degrees of freedom and the tetrads. The terms constituting the Hamiltonian of the theory can be decomposed into irreducible parts under the rotation group. Using this, we demonstrate that there are nine different classes of theories, which are distinguished by the occurrence or non-occurrence of certain primary constraints. We visualize these different classes and show that the decomposition into irreducible parts allows us to write the Hamiltonian in a common form for all nine classes, which reproduces the specific Hamiltonians of more restricted classes in which particular primary constraints appear.
\end{abstract}

\maketitle


\section{Introduction}\label{sec:intro}
General relativity (GR) is usually formulated using the Levi-Civita connection induced by a pseudo-Riemannian metric. Alternatively, one may employ other connections, such as the flat connections used in teleparallel \cite{Krssak:2015oua,Golovnev:2017dox} or symmetric teleparallel gravity \cite{BeltranJimenez:2017tkd}, in order to obtain sets of field equations equivalent to those of GR. In this work we consider teleparallel gravity, where the field variables are the 16 components of a tetrad  (or vierbein), instead of the 10 components of a metric. Nowadays it is  known that 6 components are related to local Lorentz transformations, while at most 10 encode the gravitational interaction. How many of them actually encode dynamical degrees of freedom of a  teleparallel theory of gravity is not conclusively answered in general, and to gain insight into this question is one motivation for this work.

Large varieties of teleparallel theories of gravity have been constructed \cite{Aldrovandi:2013wha,Cai:2015emx,Krssak:2018ywd}. Since the building block of these theories is the torsion of the teleparallel connection and not the curvature of the Levi-Civita connection, second order derivatives of the fundamental fields do not appear in the Lagrangians, as long as no terms with additional derivatives on the torsion are introduced, and so no Gibbons-Hawking-York boundary term is required. In this way the teleparallel formulation allows for more freedom in the construction of gravity theories with second order derivative field equations than the metric approach. Moreover, teleparallel gravity theories can be understood as gauge theories with a Yang-Mills theory like structure~\cite{BlagojevicHehl,Itin:2016nxk,Hohmann:2017duq}, which brings gravity closer to the standard model of particle physics, and might hence open a path to its unification with the other fundamental forces in physics. The other prominent reason to construct modified theories of gravity is to shed light on astrophysical observations which lack explanation within GR coupled to standard model matter only; the most famous ones being the dark matter and dark energy phenomena.

Before studying the phenomenology of modified teleparallel theories of gravity it is essential to identify those which are self-consistent, i.e.\ to understand the properties of their degrees of freedom and if they contain ghosts. This can be done best in terms of a full-fledged Hamiltonian analysis in terms of the Dirac-Bergmann algorithm for constrained Hamiltonian systems. It is known that the teleparallel equivalent of general relativity (TEGR), which yields the same dynamics and solutions for the metric defined by the tetrads as general relativity and contains no additional degrees of freedom, is self-consistent and ghost-free \cite{Okolow:2011nq,Okolow:2013lwa,Blagojevic:2000qs,Blagojevic:2000xd,Maluf:1994ji,Ferraro:2016wht,Maluf:2001rg}. The hope is that this is not the only contender of the class of healthy teleparallel theories of gravity in this sense. Because of the complexity in the calculation of the constraint algebra, the Hamiltonian analysis for modified theories of gravity is not done for all the models considered in the literature. With this work we aim to contribute to this goal.

One widely studied class of modified teleparallel theories of gravity are the $f(T)$-models. They are based on the Lagrangian $T$ employed in TEGR, and can be thought of as the teleparallel counterpart of $f(R)$-theories considered in the metric formalism. While it is known that TEGR and GR are equivalent, this is in general not true for $f(T)$ and $f(R)$ theories. The Hamiltonian analysis of $f(T)$ theories has just recently been presented \cite{Ferraro:2016wht,Ferraro:2018tpu} with the conclusion that there are three propagating degrees of freedom, which differs from previous results \cite{Chen:2014qtl,Li:2011rn}. Other, more general models are based on a Lagrangian that is a free function of the three parity even scalars that are quadratic in the torsion tensor and do not involve further fields than the tetrads~\cite{Bahamonde:2017wwk}. Their Hamilton analysis is still missing, and, due to the generality of the model, could be very involved. However, among these general models, there are the \emph{New General Relativity} (NGR) models \cite{PhysRevD.19.3524}: the most general class of teleparallel theory of gravity in four spacetime dimensions, whose Lagrangian is quadratic in the torsion tensor and contains only the tetrad and its first derivatives. This class is parametrized by three constant parameters appearing in the Lagrangian and contains TEGR for a special choice of the parameters.

Various work has been performed on NGR. Solar system constraints have been investigated~\cite{PhysRevD.19.3524} as well as the propagation and polarization modes of gravitational waves on a Minkowski spacetime background~\cite{Hohmann:2018wxu}. This analysis found that already on the linearized level, in general, NGR models predict more than two gravitational wave polarizations. However, it was also found that there exist NGR models different from TEGR with two gravitational wave polarizations. What remains open from the analysis of the linearized theory is if it differs from the full nonlinear theory. On the nonlinear level strongly coupled fields may appear, similar to what was pointed out in early attempts to formulate massive gravity theories \cite{Boulware:1973my}. A complete Hamiltonian analysis is needed in order to answer this question.

In this article we work towards the goal of a full Hamiltonian description of NGR. In particular, we derive the fully generic kinematic Hamiltonian for NGR, which is valid for any choice of the parameters appearing in the action. Further, we discuss the occurrence of primary constraints depending on the parameters of the theory. This analysis is an important cornerstone for further studies of NGR in its Hamiltonian formulation. Knowing the primary constraints, it is possible to calculate the successive Poisson brackets, and thus to derive the full constraint algebra, which implies the number of degrees of freedom of the theory. In addition it is the starting point to study the presence or absence of ghosts, and hence to test the viability of different theories within the NGR class. Further, the 3+1 Hamiltonian formalism also leads to the initial value formulation of NGR, required for numerical calculations, such as the precise prediction of gravitational wave signatures.

Hamiltonian analyses of specific theories within the NGR class besides TEGR have been studied \cite{Cheng:1988zg,Okolow:2011np}. Additionally this line of research extends to the Hamiltonian formulation of more general Poincaré gauge theories, where both torsion and curvature are present \cite{Blagojevic:1983zz,Blagojevic:1980mm}.

The main difference between the previous studies and the approach we present in this article lies in the method which is employed in order to implement the vanishing curvature of the teleparallel connection. Previous studies can mainly be divided into two groups, either assuming a vanishing spin connection (which is known as the Weitzenböck gauge)~\cite{Maluf:2001rg,Ferraro:2016wht,Ferraro:2018tpu,Okolow:2011np,Okolow:2011nq}, or an arbitrary spin connection, whose curvature is then enforced to vanish by using Lagrange multipliers in the action functional~\cite{Blagojevic:2000qs,Blagojevic:2000xd}. Here we use a different ansatz, by allowing for a non-vanishing spin connection, as mandated by the covariant formulation of teleparallel gravity~\cite{Krssak:2015oua,Golovnev:2017dox}, which is obtained explicitly by applying a local Lorentz transformation to the vanishing Weitzenböck gauge spin connection. This spin connection is flat by construction, and we will show that it enters only as a gauge degree of freedom.

The article is organized as follows: In section \ref{sec:NGRLangrangian} we present the Lagrangian for new general relativity. Then we write down the Lagrangian in 3+1 decomposition and derive its conjugate momenta, and discuss the gauge fixing, in section \ref{sec:momenta}. In section \ref{sec:Constraints} we perform a decomposition into irreducible parts and find the possible primary constraints. Finally the kinematic Hamiltonian is written down in section \ref{sec:Hamiltonian}, where we use the irreducible parts to write it in a block structure showing the most general expression. In Appendix \ref{Withoutgauge} we sketch how one can derive the Hamiltonian without fixing the gauge. Index conventions throughout this article are such that capital Latin indices $A, B, C, \ldots$ are Lorentz indices running from $0$ to $3$, Greek indices $\mu, \nu, \rho, \ldots$ are spacetime indices  running from $0$ to $3$ and small Latin indices $i, j, k, \ldots$ are spatial spacetime indices running from $1$ to $3$. A dot over a quantity always denotes derivative with respect to $x^0$  $\dot X = \partial_0 X$. The signature convention for the spacetime metric employed is $(-,+,+,+)$.

\section{The New General Relativity Lagrangian}\label{sec:NGRLangrangian}
Teleparallel theories of gravity are formulated in terms of tetrad fields $\theta^A$, their duals $e_A$ and a curvature-free spin connection $\omega^A{}_B$, which can at least locally be constructed out of local Lorentz transformations $\Lambda^A{}_B$. In local coordinates \((x^{\mu}, \mu = 0, \ldots, 3)\) on spacetime they can be expressed as
\begin{align}\label{tetrad+connection}
	\theta^A = \theta^A{}_\mu  \dd x^\mu,\quad e_A = e_A{}^\mu \partial_\mu,\quad \omega^A{}_B = \omega^A{}_{B\mu} \dd x^\mu = \Lambda^A{}_C \dd (\Lambda^{-1})^C{}_B = \Lambda^A{}_C \partial_\mu (\Lambda^{-1})^C{}_B\dd x^\mu\,,
\end{align}
and satisfy
\begin{align}\label{eq:tetradinv}
	\theta^A(e_B) = \theta^A{}_\mu e_B{}^\mu = \delta^A_B,\qquad \theta^A{}_\mu e_A{}^\nu = \delta^\nu_\mu\,.
\end{align}
Implementing the flat teleparallel spin connection in this way has the advantage that it avoids the use of Lagrange multipliers as done in \cite{Blagojevic:2000qs,Nester:2017wau}.
The spacetime metric \(g_{\mu\nu}\), which is a fundamental field in other gravity theories such as GR, here becomes a derived quantity defined by
\begin{align}
	g_{\mu\nu} = \eta_{AB}\theta^A{}_\mu \theta^B{}_\nu,\qquad g^{\mu\nu} = \eta^{AB}e_A{}^\mu e_B{}^\nu\,.
\end{align}
The fundamental tensorial ingredient from which actions for the fields are built are the first covariant derivatives of the tetrad with respect to the covariant derivative defined by the spin connection
\begin{align}\label{torsion}
	T^A = \D \theta^A = (\partial_\mu \theta^A{}_\nu + \omega^A{}_{B\mu}\theta^B{}_\nu) \dd x^\mu \wedge \dd x^\nu = \tfrac{1}{2}T^A{}_{\mu\nu}\dd x^\mu \wedge \dd x^\nu\,,
\end{align}
which is nothing but the torsion of the connection. Using the covariant derivative $\D$ in the definition of the torsion ensures a covariant transformation behavior under local Lorentz transformations of the tetrad \cite{Krssak:2015oua,Golovnev:2017dox}. Changes of index types on tensors are performed by multiplication with tetrad components, for example $T^\mu{}_{\rho\sigma} = T^A{}_{\rho\sigma} e_A{}^\mu$.

We now consider the most general Lagrange densities, in four spacetime dimensions, quadratic in torsion, which can be built from the components $T^A{}_{\mu\nu}$ of the torsion tensor and the tetrad alone, while not introducing further derivatives or parity violating terms. This class of theories can be parameterized in terms of three free parameters $c_1, c_2$ and $c_3$, and its Lagrangian is given by
\begin{align}\label{eq:LNGR}
	&L_{\text{NGR}}[\theta, \Lambda] = L_{\text{NGR}}(\theta, \partial \theta, \Lambda, \partial \Lambda)\nonumber\\
	&= |\theta| \bigg(c_1 T^\rho{}_{\mu\nu}T_{\rho}{}^{\mu\nu} + c_2T^\rho{}_{\mu\nu} T^{\nu\mu}{}_\rho + c_3 T^\rho{}_{\mu\rho}T^{\sigma\mu}{}_{\sigma}\bigg)
	= |\theta| G_{\alpha\beta}{}^{\mu\nu\rho\sigma}T^{\alpha}{}_{\mu\nu}T^\beta{}_{\rho\sigma} = |\theta| G_{AB}{}^{\mu\nu\rho\sigma}T^{A}{}_{\mu\nu}T^B{}_{\rho\sigma}\,.
\end{align}
In the last equality we introduced the convenient supermetric or constitutive tensor representation of the Lagrangian~\cite{Ferraro:2016wht,Itin:2016nxk,Hohmann:2017duq}, where below the metric must be read as a function of the tetrads \footnote{Alternatively, one may introduce the so-called axial, vector, tensor decomposition of the torsion, in which the NGR Lagrangian becomes $L = a_1 T_{\text{ax}}+a_2 T_{\text{tens}}+a_3 T_{\text{vec}}$ \cite{Bahamonde:2017wwk}. The coefficients translate as $c_1 = -\frac{1}{3}(a_1 + 2a_2)$, $c_2 = \frac{2}{3}(a_1 - a_2)$ and $c_3 = \frac{2}{3}(a_2 - a_3)$.}
\begin{align}\label{eq:G}
	G_{AB}{}^{\mu\nu\rho\sigma} =c_1 \eta_{AB}g^{\rho[\mu}g^{\nu]\sigma} - c_2 e_B^{[\mu}g^{\nu][\rho}e^{\sigma]}_A - c_3e_A^{[\mu}g^{\nu][\rho}e^{\sigma]}_B\,.
\end{align}
Teleparallel theories of gravity with the action
\begin{align}
	S[\theta,\Lambda] = \int_M L_{\text{NGR}}[\theta, \Lambda]\ \dd^4x
\end{align}
are called \emph{new general relativity} (NGR) theories of gravity \cite{PhysRevD.19.3524}. Choosing the parameters of the theory to be $c_1=\frac{1}{4}$, $c_2=\frac{1}{2}$ and $c_3=-1$ the theory reduces to TEGR~\cite{Aldrovandi:2013wha}.

\section{3+1 decomposition and conjugate momenta}\label{sec:momenta}
In order to derive the Hamilton formulation of the previously introduced NGR teleparallel theories we need to split spacetime into spatial hypersurfaces and a time direction before we derive the canonical momenta of the field variables. We introduce the $3+1$ decomposition in local coordinates $(x^0, x^i)$, where the submanifolds $x^0=const$ are the spatial hypersurfaces. As for the Hamiltonian formulation of general relativity, see, for example, the modern review \cite{Giulini:2015qha} and references therein, the metric can be decomposed into the lapse function $\alpha$, the shift vector $\beta^{i}$, and the metric on the spatial hypersurfaces~$h_{ij}$
\begin{align}
	g_{\mu\nu}=\begin{bmatrix} -\alpha^{2}+\beta^{i}\beta^{j}h_{ij} & \beta_{i} \\ \beta_{i} & h_{ij}
	\end{bmatrix}, && g^{\mu\nu}=\begin{bmatrix} -\frac{1}{\alpha^{2}} & \frac{\beta^{i}}{\alpha^{2}} \\ \frac{\beta^{i}}{\alpha^{2}} & h^{ij}-\frac{\beta^{i}\beta^{j}}{\alpha^{2}},
	\end{bmatrix}\,.
\end{align}
Spatial indices $i, j, \ldots$ are raised and lowered with the components of the spatial metric $h_{ij}$, i.e., $\beta_{i}=\beta^{j}h_{ij}$.

In the teleparallel formulation of theories of gravity we need to apply the $3+1$ decomposition to the tetrad $\theta^A = \theta^A{}_0 \dd x^0 + \theta^A{}_i \dd x^i $ and its dual $e_A = e{}_A{}^0 \partial_0 + e_A{}^i \partial_i$ instead of to the metric. They can be further expanded into lapse and shift by writing
\begin{align}
	\theta^A{}_0=\alpha \xi^{A}+\beta^{i}\theta^A{}_i\,,
\end{align}
where we introduced the components $\xi^A$ of the normal vector $n$ to the $x^0=const$ hypersurfaces in the dual tetrad basis~\cite{Okolow:2011nq}
\begin{align}
	n= \xi^A e_A,\quad \xi^{A}=-\frac{1}{6}\epsilon^{A}_{\ BCD}\theta^B{}_i\theta^C{}_j\theta^D{}_k\epsilon^{ijk}\,.
\end{align}
Lowering and raising upper-case Latin indices with the Minkowski metric $\eta_{AB}$, the $\xi^A$ satisfy
\begin{align}
	\eta_{AB}\xi^A \xi^B = \xi^{A}\xi_{A}=-1,\quad  \eta_{AB}\xi^A \theta^B{}_i = \xi_{A}\theta^A{}_i =0\,.
\end{align}
The dual tetrads and the spatial metric can be expanded into lapse, shift and spatial tetrads as
\begin{align}
	e_A{}^0=-\frac{1}{\alpha} \xi_{A},\quad e_A{}^i=\theta_A{}^i+ \xi_{A}\frac{\beta^{i}}{\alpha},\quad h_{ij} = \eta_{AB}\theta^A{}_i\theta^B{}_j\,.
\end{align}
Observe the following possible source of confusion. The spatial components of the tetrad with non-canonical index positions are defined as $\theta_A{}^i = \eta_{AB}h^{ij}\theta^B{}_j \neq e_A{}^i=\theta_A{}^i + \xi_A \frac{\beta^i}{\alpha}$. This is related to the fact that in contrast to other approaches, such as the standard calculation for the Hamiltonian of GR, we do not expand tensors into components parallel or orthogonal to the spatial hypersurfaces, but parallel to the hypersurfaces or the time direction.

Inserting these expansions into the NGR Lagrangian we obtain the $3+1$ split of the theory
\begin{align}\label{eq:NGRsplit}
\begin{split}
	L_{\text{NGR}}[\alpha, \beta^i,\theta^A{}_i,\Lambda^A{}_B]&=|\theta|\left(4 G_{AB}{}^{i0j0}T^A{}_{i0}T^B{}_{j0} + 4 G_{AB}{}^{ijk0}T^A{}_{ij}T^B{}_{k0} + G_{AB}{}^{ijkl}T^A{}_{ij}T^B{}_{kl}\right)\\
	&=\frac{\sqrt{h}}{2\alpha}T^{A}{}_{i0}T^{B}{}_{j0}M^i{}_A{}^j{}_B
	\\&+\frac{\sqrt{h}}{\alpha}T^{A}{}_{i0}T^{B}{}_{kl}\left[M^i{}_A{}^l{}_B\beta^{k}+2\alpha h^{il} (c_{2}\xi_{B}\theta_A{}^k+c_{3}\xi_{A}\theta_B{}^k) \right]
	\\&+ \frac{\sqrt{h}}{\alpha}T^{A}{}_{ij}T^{B}{}_{kl}\beta^{i}\left[\tfrac{1}{2}M^j{}_A{}^l{}_B \beta^{k}+2 \alpha h^{jl} (c_{2} \xi_{B}\theta_A{}^k+ c_{3}\xi_{A}\theta_B{}^k) \right]
	\\&+\alpha \sqrt{h}\cdot {}^{3}\mathbb{T}\,.
\end{split}
\end{align}
The matrix $M^i{}_A{}^j{}_B$ is a map from $3\times4$ matrices to their duals, i.e. $4\times3$ matrices, and will play an important role when we express the velocities of the tetrads in terms of the canonical momenta and vice versa. It can be written in the form
\begin{align}\label{eq:Mdef}
\begin{split}
	M^i{}_A{}^j{}_B
	&= 8\alpha^{2}G_{AB}{}^{i0j0}\\
	&=-2(2 c_{1}h^{ij}\eta_{AB} - (c_{2}+c_{3})\xi_{A}\xi_{B}h^{ij} + c_{2}\theta_A{}^j\theta_B{}^i + c_{3}\theta_A{}^i\theta_B{}^j)\,.
\end{split}
\end{align}
The purely intrinsic torsion scalar on the $x^0=const$ hypersurface is given by
\begin{align}
\begin{split}
	{}^{3}\mathbb{T}\equiv c_{1}\eta_{AB}T^A{}_{ij}T^B{}_{kl}h^{ik}h^{jl}+c_{2}\theta_A{}^i \theta_B{}^j T^A{}_{kj}T^B{}_{li}h^{kl}+c_{3}\theta_A{}^i\theta_B{}^j h^{kl}T^A{}_{ki}T^B{}_{lj} = H_{AB}{}^{ijkl}T^A{}_{ij}T^B{}_{kl}\,,
\end{split}
\end{align}
where the spatial supermetric is
\begin{align}
	H_{AB}{}^{ijkl} = c_1 \eta_{AB}h^{k[i}h^{j]l} - c_2 \theta_B{}^{[i}h^{j][k}\theta^{l]}{}_A - c_3\theta_A{}^{[i}h^{j][k}\theta^{l]}{}_B\,.
\end{align}

In the $3+1$ decomposed form \eqref{eq:NGRsplit} it is not difficult to derive the canonical momenta of the tetrads $\theta^A{}_\mu$ and the Lorentz transformations $\Lambda^A{}_B$ which generate the spin connection. Time derivatives on the variables of the theory only appear in torsion terms $T^A{}_{0i}$ and never act on $\theta^A{}_0$, due to the antisymmetry of the torsion tensor in its lower indices, nor on the lapse $\alpha$ and the shift $\beta$. Hence the canonical momenta of lapse and shift are, not surprisingly, all identically zero,
\begin{align}\label{eq:pilapseshift}
	\pi_\alpha = \frac{\partial L_{\text{NGR}}}{\partial \dot\alpha} = 0,\quad \pi_{\beta^i} = \frac{\partial L_{\text{NGR}}}{\partial \dot\beta^i} = 0\,.
\end{align}
The canonical momenta of the spatial tetrad components are given by
\begin{align}
\begin{split}
\label{conj}
	\frac{\alpha}{\sqrt{h}} \pi_A{}^i= \frac{\alpha}{\sqrt{h}} \frac{\partial L_{\text{NGR}}}{\partial \dot \theta^A{}_i} = T^B{}_{0j}M^i{}_A{}^j{}_B + T^B{}_{kl}\left[M^i{}_A{}^k{}_B\beta^{l}+2\alpha h^{ik}( c_{2}\xi_{B}\theta_A{}^l+c_{3}\xi_{A}\theta_B{}^l) \right]\,,
\end{split}
\end{align}
while the momenta for the connection generating Lorentz transformations turn out to be completely determined from the momenta of the tetrad.



To see this first observe that the Lorentz group is six dimensional and therefore not all components of the $\Lambda^A{}_B$ are independent of each other. To reflect this during the derivation of the corresponding momenta we introduce the auxiliary antisymmetric field $a_{AB}$ in the following way:
\begin{align}
\label{Defa}
a_{AB}:=\eta_{AC}\omega^C{}_{B0} = \eta_{C[A}\Lambda^C{}_{|D|} \dot{(\Lambda^{-1})}^D{}_{B]}
\Leftrightarrow
\dot \Lambda^A{}_B=a_{MN}\eta^{A[N} \Lambda^{M]}{}_B \,.
\end{align}
The independent components of the momenta of the Lorentz matrices are then given by
\begin{align}
	\label{conjugatemomentarelation}
	\hat{\pi}^{AB} = \frac{\partial L_{\text{NGR}}}{\partial a_{AB}}\,,
\end{align}
and satisfy
\begin{align}
		\label{conjugatemomentaauxillaryrel}
		\hat{\pi}^{AB}&=- \pi_C{}^i\eta^{C[B}\theta^{A]}{}_i \,,
\end{align}
which can easily be realized from
\begin{align}
	\frac{\partial L_{\text{NGR}}}{\partial a_{MN}}
	&=\frac{\partial L_{\text{NGR}}}{\partial \dot \Lambda^A{}_B} \frac{\partial \dot \Lambda^A{}_B}{\partial a_{MN}}
	= \frac{\partial L_{\text{NGR}}}{\partial T^C{}_{0k}} \frac{\partial T^C{}_{0k}}{\partial \dot \Lambda^A{}_B} \frac{\partial \dot \Lambda^A{}_B}{\partial a_{MN}}
	= -\frac{\partial L_{\text{NGR}}}{\partial T^C{}_{0k}} \frac{\partial T^C{}_{0k}}{\partial \dot \theta^A{}_i} \left[\theta^D{}_i(\Lambda^{-1})^B{}_D\right] \frac{\partial \dot \Lambda^A{}_B}{\partial a_{MN}}\\
	&= -\frac{\partial L_{\text{NGR}}}{\partial \dot \theta^A{}_i}  \left[\theta^D{}_i(\Lambda^{-1})^B{}_D\right]\eta^{A[N}\Lambda^{M]}{}_B\,.
\end{align}

The fact that the momenta $\hat \pi$ are not independent of the momenta $\pi$ demonstrates that the $\Lambda^A{}_B$ are not independent, but only gauge degrees of freedom.

In the following, we introduce new field variables $(\tilde \alpha, \tilde \beta^i, \tilde \theta^A{}_i, \tilde \Lambda^A{}_B)$, where $\tilde{\theta}{}^A{}_i(\theta,\Lambda) := \theta^B{}_i (\Lambda^{-1})^A{}_B$ is the so-called Weitzenb\"ock tetrad and all other fields are not changed: $\tilde \alpha=\alpha,\ \tilde \beta^i = \beta^i$ and $\tilde \Lambda^A{}_B = \Lambda^A{}_B$. Using the inverse of this definition $\theta^B{}_i = \tilde{\theta}{}^A{}_i \Lambda^B{}_A$ to express the Lagrangian \eqref{eq:LNGR} in terms of the Weitzenb\"ock tetrad yields that $\tilde L_{\text{NGR}}[\alpha, \beta^i,\tilde \theta^A{}_i, \Lambda^A{}_B] := L_{\text{NGR}}[\alpha, \beta^i,\theta^A{}_i(\tilde \theta, \Lambda),\Lambda^A{}_B]$ is independent of~$\Lambda$, respectively, $\tilde \Lambda$. The $\alpha$ and $\beta^i$ momenta are not affected by this field redefinition at all. For the momenta in the new frame we find the transformation behavior
\begin{align}\label{eq:Ptrafo1}
\tilde{\pi}_A{}^i
= \frac{\partial \tilde L_{\text{NGR}}}{\partial\dot{\tilde{ \theta}}^A{}_i}
= \pi_B{}^i\Lambda^B{}_A\,,\quad
\hat {\tilde \pi}^{MN}
= \frac{\partial \tilde L_{\text{NGR}}}{\partial a_{MN}}
=  \pi_A{}^j  \eta^{A[N}\theta^{M]}{}_j + \hat \pi^{MN} \,,
\end{align}
with inverse transformation
\begin{align}\label{eq:Ptrafo3}
\pi_A{}^i = \tilde \pi_B{}^i (\Lambda^{-1}){}^B{}_A,\quad \hat \pi^{MN} = \hat {\tilde \pi}^{MN} - \tilde \pi_B{}^j (\Lambda^{-1}){}^B{}_A  \eta^{A[N}\Lambda^{M]}{}_C \tilde \theta^C{}_j \,.
\end{align}
Applying the constraint \eqref{conjugatemomentaauxillaryrel} to the second part of the transformation \eqref{eq:Ptrafo1} shows that in the Weitzenb\"ock gauge the momenta of the Lorentz transformations all vanish, $\hat {\tilde \pi}^{AB} = 0$.

This reproduces the well-known fact that in teleparallel gravity the spin connection represents pure gauge degrees of freedom~\cite{Krssak:2015oua,Golovnev:2017dox}. Therefore, without loss of generality, we can set the spin connection coefficients to zero and work in the so-called Weitzenböck gauge, in which the connection coefficients of the spin connection vanish identically.

The Hamiltonian in the Weitzenb\"ock gauge is then given by the Legendre transform of the the Lagrangian where we have to add the primary constraints we already discovered, \eqref{eq:pilapseshift} and \eqref{conjugatemomentaauxillaryrel} with Lagrange multipliers ${}^{\tilde \alpha}\lambda$, ${}^{\tilde \beta}\lambda^i$ and~${}^{\hat \pi}\lambda$
\begin{align}\label{eq:HNGR}
\begin{split}
	\tilde H_{\text{NGR}}[{}^{\tilde \alpha}\lambda, {}^{\tilde \beta}\lambda^i, {}^{\tilde\pi}\lambda_{AB},\tilde \alpha,\tilde\pi_\alpha, \tilde \beta^i,\tilde\pi_{\beta^i},\tilde \theta^A{}_i,\hat \pi_A{}^i, \tilde \Lambda^A{}_B, \hat {\tilde \pi}^{AB} ] 
	&= \tilde\pi_\alpha\dot{\tilde\alpha} + \tilde\pi_{\beta^i}\dot{\tilde\beta}^i + \tilde \pi_A{}^i \dot {\tilde \theta}^A{}_i + \hat {\tilde \pi}^{AB} \tilde a_{AB} \\
	&+ {}^{\tilde \alpha}\lambda \tilde\pi_\alpha + {}^{\tilde \beta}\lambda^i \tilde\pi_{\beta^i} + {}^{\hat {\tilde{\pi}}}\lambda_{AB} \hat {\tilde \pi}^{AB}
	 - \tilde L_{NGR}[\tilde \alpha, \tilde \beta^i,\tilde \theta^{}_i, \tilde \Lambda]\,.
\end{split}
\end{align}
The term $\hat {\tilde \pi}^{AB} \tilde a_{AB}$ is identical to the term one would use naively in terms of the canonical variables $\frac{\tilde \partial L_{\text{NGR}}}{\partial \dot{\Lambda}^A{}_B}\dot \Lambda^A{}_B$, as can easily be seen from the definition of the auxiliary variable $a_{AB}$ in \eqref{Defa}. As mentioned $\tilde \alpha = \alpha$, $\tilde \beta^i = \beta^i$ and $\tilde \Lambda^A{}_B = \Lambda^A{}_B$ and $\tilde L_{\text{NGR}}[\tilde \alpha, \tilde \beta^i,\tilde \theta^{}_i, \tilde \Lambda]$ is independent of $\Lambda$. Therefore, on shell, where the constraint $\hat {\tilde \pi}^{AB}=0$ is implemented, the gauge fixed Hamiltonian does neither depend on $\Lambda$ nor on $\hat {\tilde \pi}^{AB}$. Moreover the evolution of the constraints is preserved since their Poisson bracket with the Hamiltonian vanishes $\{\tilde\pi_\alpha,\tilde H\}\approx 0$, $\{\tilde\pi_{\beta^i},\tilde H\}\approx 0$,  $\{\hat {\tilde \pi}^{AB},\tilde H\} \approx 0$ on the constraint surface $\tilde\pi_\alpha =\tilde\pi_{\beta^i} =\hat {\tilde \pi}^{AB} =0$.

These findings on the level of canonical momenta demonstrate that we do not need to include the variables $\tilde\pi_\alpha,\ \tilde\pi_{\beta^i},\ \Lambda$ and $\hat \pi$ in the Hamiltonian and again justify the approach in \cite{Maluf:2000ag}. In the following we will work in the Weitzenb\"ock gauge and omit the tilde from $\tilde{\theta},\tilde{\pi},\hat{\tilde{\pi}}$ for readability.

\section{Inverting the momentum-velocity relation}\label{sec:Constraints}
One essential step in the reformulation of a physical field theory from its Lagrangian to its Hamiltonian description is to invert the relation between the momenta and the velocities, to express the latter in terms of the former. For NGR this amounts to inverting the equation \eqref{conj}. To do so we rewrite the equation as a linear map from the space of $4 \times 3$ matrices to the space of $3 \times 4$ matrices
\begin{align}
\label{invertvelocity}
S_A{}^i= M^i{}_A{}^j{}_B \dot{\theta}^B{}_j\,,
\end{align}
with a source term $S_A{}^i$, which only depends on the momenta, the fields and their spatial derivatives,
\begin{align}\label{eq:source}
\begin{split}
S_A{}^i[\alpha,\beta,\theta^A{}_i,\pi_A{}^i]=\frac{\alpha}{\sqrt{h}} \pi_A{}^i + \big[ D_{k} \left(\alpha \xi^{B} + \beta^{m}\theta^B{}_m \right) - T^B{}_{kl}\beta^{l} \big] M^i{}_A{}^k{}_B
- 2 \alpha T^B{}_{kl} h^{ik}( c_{2}\xi_{B}\theta_A{}^l+c_{3}\xi_{A}\theta_B{}^l),
\end{split}
\end{align}
where $D_{i}$ is the Levi-Civita covariant derivative of the hypersurface metric \(h_{ij}\). By inverting this equation we can re-express the field velocities in terms of the canonical variables: the fields themselves and their momenta.

To explicitly invert equation \eqref{invertvelocity} we decompose the velocities of the spatial tetrads into irreducible parts with respect to the rotation group. It turns out that in this decomposition the matrix $M$ has a block diagonal structure which can be inverted block by block. Since for certain combinations of the $c_{1},c_{2},c_{3}$ parameters of the theory some blocks become identically zero, we employ the Moore-Penrose pseudoinverse of a matrix \cite{Ferraro:2016wht} to display the inverse in a closed form for all choices of the parameters. This then carries over when we display the Hamiltonian.

The irreducible decomposition with respect to the rotation group amounts in defining a vectorial ($\mathcal{V}$), antisymmetric ($\mathcal{A}$), symmetric trace-free ($\mathcal{S}$), and trace ($\mathcal{T}$) part of the tetrad velocities and their momenta:
\begin{align}\label{veldec}
\dot{\theta}^A{}_i
&={}^{\mathcal{V}}\dot{\theta}_{i}\ \xi^{A}+{}^{\mathcal{A}}\dot{\theta}_{ji}\ h^{kj}\theta^A{}_k+{}^{\mathcal{S}}\dot{\theta}_{ji}\ h^{kj}\theta^A{}_k+{}^{\mathcal{T}}\ \dot{\theta}\ \theta^A{}_i,\\
\pi_A{}^i
&={}^{\mathcal{V}}\pi^{i}\ \xi_{A}+{}^{\mathcal{A}}\pi^{ji}\ h_{kj}\theta_A{} ^k+{}^{\mathcal{S}}\pi^{ji}\ h_{kj}\theta_A{}^k+{}^{\mathcal{T}}\pi\ \theta_A{}^i\,.
\end{align}
Decomposing $S_A{}^i$ into the same irreducible parts and using the explicit form of $M$, see equation \eqref{eq:Mdef}, yields
\begin{align}\label{eq:VS}
\begin{split}
	{}^\mathcal{V}S^i
	&=- \xi^AS_A{}^i \\
	&={}^{\mathcal{V}}\pi^{i} \frac{\alpha}{\sqrt{h}}  - 2 \alpha c_{3}T^B{}_{kl} h^{ik}\theta_B{}^l + 2 (2c_{1}+c_{2}+c_{3})\big[D_{k}\big(\alpha \xi^{B}+\beta^{m}\theta^B{}_m)- T^B{}_{kl}\beta^{l} \big]\xi_{B}h^{ik} \\
	&= -2\ {}^{V}\dot{\theta}_{j}\ h^{ij}(2c_{1}+c_{2}+c_{3})\,,
\end{split}
\end{align}
for the vector part,
\begin{align}\label{eq:AS}
\begin{split}
	{}^\mathcal{A}S^{mp}
	&= \theta^A{}_i h^{i[m}S_A{}^{p]} \\
	&={}^\mathcal{A}\pi^{mp} \frac{\alpha}{\sqrt{h}} - 2 \alpha c_{2} h^{lm}h^{pk}T^B{}_{kl} \xi_{B}  - 2( 2 c_1 - c_2) \big[D_{k}\left( \alpha \xi^{B}+\beta^{s}\theta^B{}_s\right)-	T^B{}_{kl} \beta^l\big] \theta_B{}^{[m}h^{p]k} \\
	&=- 2\ {}^\mathcal{A}\dot{\theta}^{mp}\ (2c_{1} - c_2)
\end{split}
\end{align}
for the antisymmetric part,
\begin{align}\label{eq:SS}
\begin{split}
	{}^\mathcal{S}S^{mp}
	&= \theta^A{}_q h^{q(m}S_A{}^{p)}  - \tfrac{1}{3} \theta^A{}_i S_A{}^i h^{mp} \\
	&={}^\mathcal{S}\pi^{mp} \frac{\alpha}{\sqrt{h}} - 2(2c_{1}+c_{2}) \big[ D_{k}\left( \alpha \xi^{B}+\beta^{s}\theta^B{}_s \right) -
	T^B{}_{kl}\beta^l \big] \left(\theta_B{}^{(m}h^{p)k}-\tfrac{1}{3}h^{pm}\theta_B{}^k\right)\\
	&=- 2\ {}^{\mathcal{S}}\dot{\theta}^{mp} (2 c_1 +c_2)
\end{split}
\end{align}
for the trace-free symmetric part, and
\begin{align}\label{eq:TS}
\begin{split}
	{}^\mathcal{T}S
	&= \tfrac{1}{3}   \theta^A{}_iS_A{}^i \\
	&={}^\mathcal{T}\pi \frac{\alpha}{\sqrt{h}} -\tfrac{2}{3}(2c_{1}+c_{2}+3c_{3}) \big[ D_{k}\left(\alpha \xi^{B}\beta^{m}\theta^B{}_m \right)
	- T^B{}_{kl}\beta^{l}\Big]\theta_B{}^k\\
	&= -2\ {}^{T}\dot{\theta}\ (2c_{1}+c_{2}+3c_{3})
\end{split}
\end{align}
for the trace part.

These equations are easily solved for the velocities in terms of their dual momenta in case the coefficients
\begin{align}
	A_{\mathcal{V}} = 2c_{1}+c_{2}+c_{3},\quad A_{\mathcal{A}} = 2c_{1}-c_{2},\quad A_{\mathcal{S}} = 2c_{1}+c_{2} \textrm{ and } A_\mathcal{T}=2c_{1}+c_{2}+3c_{3}\,,
\end{align}
are all non-vanishing. In case one or more of these coefficients vanish they induce primary constraints
\begin{alignat}{4}
	A_{\mathcal{V}} &= 0 &\quad&\Rightarrow & \quad{}^\mathcal{V}C^i &:= \frac{{}^{\mathcal{V}}\pi^{i}}{\sqrt{h}} - 2  c_{3}T^B{}_{kl} h^{ik}\theta_B{}^l &&= 0 \label{eq:C1}\\
	A_{\mathcal{A}} &= 0 &\quad&\Rightarrow & \quad{}^\mathcal{A}C^{ij} &:= \frac{{}^\mathcal{A}\pi^{ij}}{\sqrt{h}} - 2  c_{2} h^{li}h^{jk}T^B{}_{kl} \xi_{B} &&= 0 \label{eq:C2}\\
	A_{\mathcal{S}} &= 0 &\quad&\Rightarrow & \quad{}^\mathcal{S}C^{ij} &:=  \frac{{}^\mathcal{S}\pi^{ij}}{\sqrt{h}} &&= 0 \label{eq:C3}\\
	A_{\mathcal{T}} &= 0 &\quad&\Rightarrow & \quad{}^\mathcal{T}C &:=  \frac{{}^\mathcal{T}\pi}{\sqrt{h}} &&= 0 \label{eq:C4}\,.
\end{alignat}
Observe that ${}^VC^i$ correspond to 3 constraints, ${}^AC^{mp}$ to 3 (since it is antisymmetric in its indices), ${}^SC^{mp}$ to 5 (since it is symmetric in its indices, but does not contain the trace part), and ${}^TC$ corresponds to 1 constraint. For any choice of the parameters $c_1,\ c_2,\ c_3$ we either can invert the appearing velocities of the tetrads in terms of the tetrads and their momenta, or we obtain a constraint from the Lagrangian, which must be implemented in the Hamiltonian later by a Lagrange multiplier.

The Moore-Penrose pseudoinverse of the matrix $M$ in the irreducible decomposition of the rotation group we employed is given by the inverse of the separate blocks if the coefficient in front of the block $A_{\mathcal{V}},A_{\mathcal{A}},A_{\mathcal{S}}$ or $A_{\mathcal{T}}$ is non-vanishing. In case one of the coefficients is vanishing the block in the inverse matrix is simply a block of zeros. For completeness we display $M$ and its Moore-Penrose pseudoinverse explicitly. Expanding $M$ itself into the irreducible parts basis
\begin{align}
	M^i{}_A{}^j{}_B = {}^\mathcal{V}M^{ij}\ \xi_A \xi_B + {}^\mathcal{A} M^{[ir][js]}\ \theta^C{}_r \eta_{AC}\theta^D{}_s \eta_{BD} + {}^\mathcal{S} M^{(ir)(js)}\ \theta^C{}_r \eta_{AC}\theta^D{}_s \eta_{BD} + {}^\mathcal{T} M\ \theta_A{}^i \theta_B{}^j
\end{align}
yields
\begin{align}
\begin{split}
	M^i{}_A{}^j{}_B
	&= 2 A_{\mathcal{V}}\ \xi_A \xi_B h^{ij} - 2 A_{\mathcal{A}}\ h^{i[j}h^{s]r}\theta^C{}_r \eta_{AC}\theta^D{}_s \eta_{BD}\\
	&- 2 A_{\mathcal{S}}\ (h^{i(j}h^{s)r}- \tfrac{1}{3}h^{ir}h^{js})\theta^C{}_r \eta_{AC}\theta^D{}_s \eta_{BD} - \frac{2}{3} A_{\mathcal{T}}\ \theta_A{}^i \theta_B{}^j\,.
\end{split}
\end{align}
By using the identity $\eta^{AB}+\xi^A\xi^B = \theta^A{}_i \theta^B{}_j h^{ij}$ one may check that this representation of $M$ is indeed identical to its definition \eqref{eq:Mdef}. Its pseudoinverse is
\begin{align}
\begin{split}
	\label{eq:MinvRed}
	\left(M^{-1}\right)^{A \ C}_{\ i \ k}&= \frac{1}{2}B_{\mathcal{V}} \xi^{A}\xi^{C}h_{ik} -\frac{1}{2} B_{\mathcal{A}}  h^{r[s}h^{m]n} h_{kr}h_{si}\theta^A{}_m\theta^C{}_n
	\\& -\frac{1}{2} B_{\mathcal{S}} \left( h^{r(s}h^{m)n} - \tfrac{1}{3} h^{sm} h^{nr} \right) h_{kr}h_{si}\theta^A{}_m\theta^C{}_n-\frac{1}{6} B_{\mathcal{T}} \theta^A{}_i\theta^C{}_k,
\end{split}
\end{align}
where the different blocks are implemented by defining ($I={\mathcal{V}},{\mathcal{A}},{\mathcal{S}},{\mathcal{T}}$)
$
	B_I=
	\begin{cases}
	0 &\textrm{ for } A_I = 0\\
	\frac{1}{A_I} &\textrm{ for } A_I \neq 0\,.
	\end{cases}
$

\section{The NGR Hamiltonian}\label{sec:Hamiltonian}
To obtain the Hamiltonian from the Lagrangian we use its definition as Legendre transform omitting the variables $\Lambda$ and $\hat \pi$, as discussed below equation \eqref{eq:HNGR}. We display the dependencies on the remaining variables explicitly for clarification, and the square brackets shall indicate that the function may depend on the spatial derivatives of the fields,
\begin{align}
	H[\alpha, \beta^i, \theta^A{}_j, \pi_A{}^k] = \dot \theta^A{}_i[\alpha, \beta^i, \theta^A{}_j, \pi_A{}^k] \pi_A{}^i - L[\alpha, \beta^i, \theta^A{}_j, \dot \theta^A{}_k[\alpha, \beta^r, \theta^A{}_s, \pi_A{}^m]]
\end{align}
We will suppress these dependencies in the brackets from now on for the sake of readability. Moreover, we comment on how to remove the gauge fixing, i.e.\ how to reintroduce the dependence on $\Lambda$ and $\hat \pi$ at the end of this section. A sketch on how the calculations would be carried out without gauge fixing is made in Appendix \ref*{Withoutgauge}.

To derive the Hamiltonian explicitly we can first use the source expression $S$, defined in equation \eqref{eq:source}, to simplify the Lagrangian. This can be done by expanding the $T^A{}_{i0}$ terms in equation \eqref{eq:LNGR} into the time derivatives of the tetrad and combining them with the $M$ matrices to the source term whenever possible. By their definition, they can then be expanded in terms of the momenta and spatial derivatives acting on the fields.  As an intermediate result the Hamiltonian becomes
 \begin{align}\label{eq:protoH}
 	\begin{split}
 	H
 	&= \frac{1}{2}\dot{\theta}^A{}_i\pi_A{}^i-\sqrt{h}T^B{}_{jk}\dot{\theta}^A{}_i h^{ij}\left[c_{2}\xi_{B}\theta_A{}^k + c_{3}\xi_{A}\theta_B{}^k \right] \\
 	&+\frac{1}{2}\pi_A{}^iD_{i}\left(\alpha \xi^{A}+\beta^{j}\theta^A{}_j \right)+\sqrt{h}T^B{}_{ kl}D_{i}\left( \alpha \xi^{A}+\beta^{j}\theta^A{}_j\right) h^{ik}\left[c_{2}\xi_{B}\theta_A{}^l+c_{3}\xi_{A}\theta_B{}^l \right]\\
 	&-\frac{1}{2}\pi_B{}^jT^B{}_{jk}\beta^{k} -\sqrt{h}T^A{}_{ij}T^B{}_{kl}\beta^{k}h^{il} \left[ c_{2}\xi_{A}\theta_B{}^j+c_{3}\xi_{B}\theta_A{}^j\right] -\alpha \sqrt{h}\cdot{}^{3}\mathbb{T} 
 	\end{split}
 \end{align}
To eliminate the remaining velocities we expand them into the ${\mathcal{V}},{\mathcal{A}},{\mathcal{S}},{\mathcal{T}}$ decomposition, we introduced in the previous section, and replace them according to equations \eqref{eq:VS} to \eqref{eq:TS}.

Expanding the first term in the irreducible decomposition yields
\begin{align}\label{eq:dthetapi}
	 \dot \theta^A{}_i \pi_A{}^i
	  &= - {}^{\mathcal{V}}\dot{\theta}_{i}\ {}^{\mathcal{V}}\pi^i + {}^{\mathcal{A}}\dot{\theta}_{ji}\ {}^{\mathcal{A}}\pi^{ji} +{}^{\mathcal{S}}\dot{\theta}_{ji}\ {}^{\mathcal{S}}\pi^{ji} + 3 {}^{\mathcal{T}}\dot{\theta}\ {}^{\mathcal{T}}\pi\\
	 &= \alpha \Bigg(\frac{{}^\mathcal{V}C_i\ {}^\mathcal{V}\pi^i}{2 A_\mathcal{V}} - \frac{{}^\mathcal{A}C_{ij}\ {}^\mathcal{A}\pi^{ij}}{2 A_\mathcal{A}} - \frac{{}^\mathcal{S}C_{ij}\ {}^\mathcal{S}\pi^{ij}}{2 A_\mathcal{S}} - \frac{3 {}^\mathcal{T}C\ {}^\mathcal{T}\pi}{2 A_\mathcal{T}}\Bigg) + \pi_A{}^i D_i\left(\alpha \xi^{A}+\beta^{m}\theta^A{}_m \right) - \pi_A{}^i T^A{}_{im}\beta^m  \,.
\end{align}
while for the second we find
\begin{align}\label{eq:dthetaT}
\begin{split}
	\sqrt{h}T^B{}_{jk}\dot{\theta}^A{}_i h^{ij}\left[c_{2}\xi_{B}\theta_A{}^k + c_{3}\xi_{A}\theta_B{}^k \right]
	&= c_2 T^B{}_{jk}\ {}^\mathcal{A}\dot{\theta}_{mi}\ h^{km} h^{ij} - c_3 T^B{}_{jk}\ {}^\mathcal{V}\dot\theta_i\ h^{ij} \theta_B{}^k\\
	&= \frac{\alpha}{2 A_{\mathcal{A}}}c_2 \xi_B T^B{}_{jk}\ {}^\mathcal{A}C^{jk} + \frac{\alpha}{2 A_{\mathcal{V}}}c_3 \theta_B{}^k T^B{}_{jk}\ {}^\mathcal{V}C^{j} \\
	&- [D_i\left(\alpha \xi^{C}+\beta^{m}\theta^C{}_m \right) - T^C{}_{im}\beta^m]T^B{}_{jk}h^{ki}[c_2 \xi_B \theta_C{}^j + c_3 \xi_C \theta_B{}^j]\,.
\end{split}
\end{align}
Inserting the expressions \eqref{eq:dthetapi} and \eqref{eq:dthetaT} into equation \eqref{eq:protoH} finally yields the kinematic Hamilton density of the NGR teleparallel theories of gravity,

\begin{align}\label{eq:Hkin}
\begin{split}
	H
	&=  \alpha \sqrt{h}\Bigg(\frac{{}^\mathcal{V}C_i\ {}^\mathcal{V}C^i}{4 A_\mathcal{V}} - \frac{{}^\mathcal{A}C_{ij}\ {}^\mathcal{A}C^{ij}}{4 A_\mathcal{A}} - \frac{{}^\mathcal{S}C_{ij}\ {}^\mathcal{S}C^{ij}}{4 A_\mathcal{S}} - \frac{3 {}^\mathcal{T}C\ {}^\mathcal{T}C}{4 A_\mathcal{T}} - {}^{3}\mathbb{T}  - \frac{\xi^A D_i\pi_A{}^i}{\sqrt{h}}\Bigg)  - \beta^k(T^A{}_{jk}\pi_A{}^j + \theta^A{}_k D_{i}\pi_A{}^i)\\
	& + D_i[\pi_A{}^i(\alpha \xi^{A}+\beta^{j}\theta^A{}_j )]
\end{split}
\end{align}
which we here display in terms of the constraints \eqref{eq:C1} to \eqref{eq:C4}, as this is the most convenient expression. Observe that, even though we use the irreducible $\mathcal{V},{\mathcal{A}},{\mathcal{S}},\mathcal{T}$ decomposition of the fields to display the Hamiltonian, since in this form the dependence on the parameters $c_i$ becomes most clear, the canonical variables on which the Hamiltonian depends are $\{\alpha, \beta^i, \theta^A{}_j, \pi_A{}^k\}$. As in general relativity we immediately see that we deal with a pure constraint Hamiltonian up to boundary terms. Lapse $\alpha$ and shift $\beta$ have vanishing momenta, $\pi_\alpha = 0$ and $\pi_{\beta_i} = 0$, and appear only as Lagrange multipliers.
To obtain the dynamically equivalent Hamiltonian to the Lagrangian \eqref{eq:LNGR} we need to add possible further nontrivial constraints via Lagrange multipliers. To find all constraints it is necessary to calculate the Poisson brackets between all primary constraints, check if they are first class, and, in case they are not, add possible secondary constraints. This algorithm has to be continued until a closed constraint algebra is obtained \cite{Dirac}.

\begin{figure}[htb]
\includegraphics[width=0.8\textwidth]{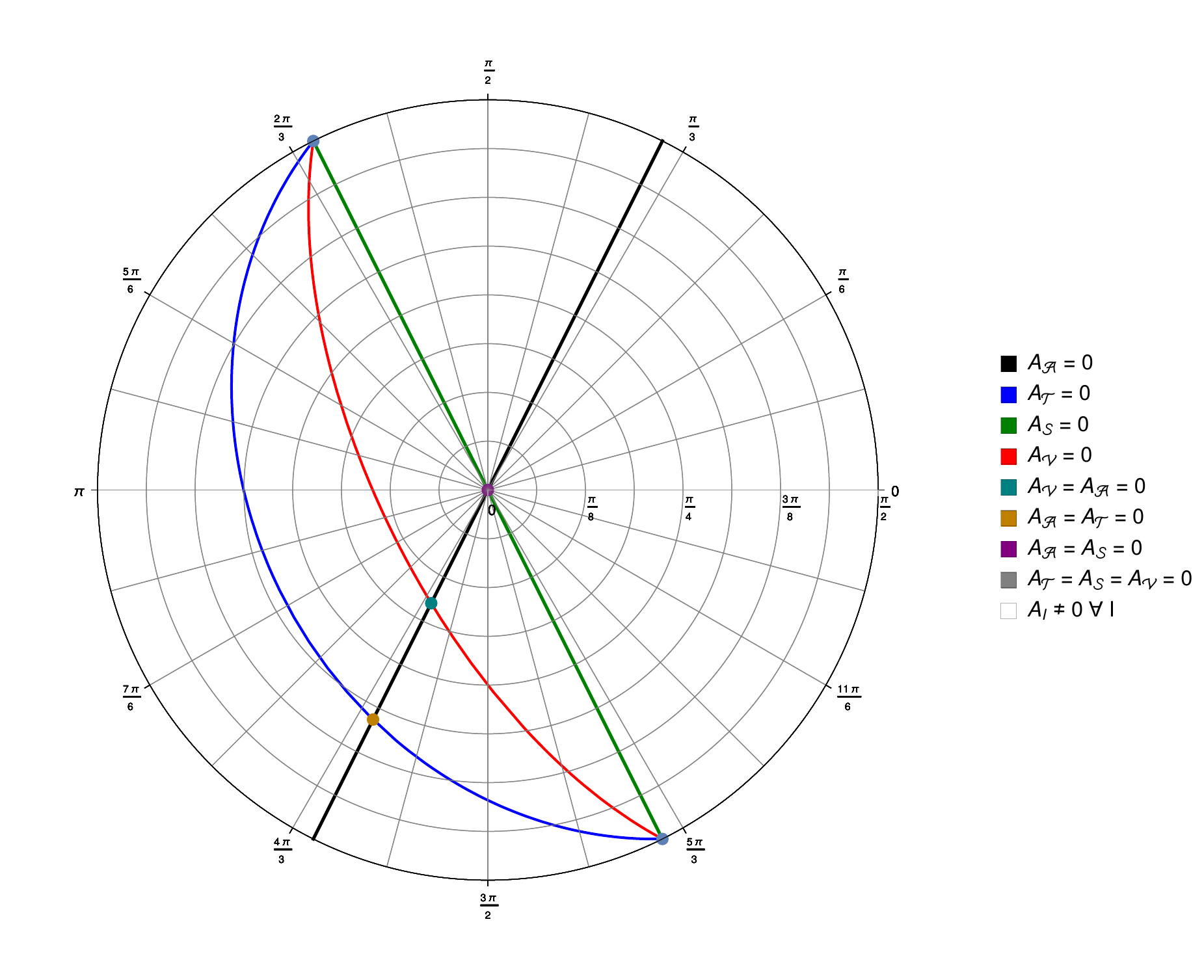}
\caption{(Color online.) Visualization of the parameter space of new general relativity, colored by the occurrences of primary constraints. The radial axis shows the zenith angle \(\theta\), while the (circular) polar axis shows the azimuth angle \(\phi\), following the definition~\eqref{eqn:polcoord}.}
\label{fig:conplot}
\end{figure}

From our analysis in section \ref{sec:Constraints} we conclude that the NGR theories of gravity decay into nine subclasses depending on the choice of the parameters $c_1$ , $c_2$ and $c_3$, which correspond to the appearance of different primary class constraints, in addition to the lapse and shift constraints arising from the diffeomorphism invariance of the action. We have visualized these classes in figure~\ref{fig:conplot}, which we constructed as follows. We started from the assumption that at least one of the parameters \(c_1, c_2, c_3\) is non-vanishing, since otherwise the Lagrangian would be trivial, and introduced normalized parameters
\begin{equation}
\tilde{c}_i = \frac{c_i}{\sqrt{c_1^2 + c_2^2 + c_3^2}}
\end{equation}
for \(i = 1, 2, 3\). One easily checks that the constraint classes we found only depend on these normalized parameters. We then introduced polar coordinates \((\theta, \phi)\) on the unit sphere to express the parameters as
\begin{equation}\label{eqn:polcoord}
\tilde{c}_1 = \sin\theta\cos\phi\,, \quad
\tilde{c}_2 = \sin\theta\sin\phi\,, \quad
\tilde{c}_3 = \cos\theta\,.
\end{equation}
Since the same constraints appear for antipodal points on the parameter sphere, we restrict ourselves to the hemisphere \(\tilde{c}_3 \geq 0\), and hence \(0 \leq \theta \leq \frac{\pi}{2}\); this is equivalent to identifying antipodal points on the sphere and working with the projective sphere instead, provided that we also identify antipodal points on the equator \(\tilde{c}_3 = 0\). We then considered \((\theta, \phi)\) as polar coordinates on the plane in order to draw the diagram shown in figure~\ref{fig:conplot}. Note that antipodal points on the perimeter, such as the two gray points for the most constrained case, are identified with each other, since they describe the same class of theories. To summarize, we find the following constraints:

\begin{center}
	\begin{tabular}{|c|c|c|}
		\hline
		Theory & Constraints & Location in figure~\ref{fig:conplot} \\ \hline
		$A_{I}\neq 0 \ \forall I\in \{ \mathcal{V},{\mathcal{A}},{\mathcal{S}},\mathcal{T} \}$ &  No constraints & white area
		\\ \hline
		$A_{\mathcal{V}}=0$ &  ${}^{\mathcal{V}}C_{i}=0$ & red line \\ \hline
		$A_{\mathcal{A}}=0$ &   ${}^{\mathcal{A}}C_{ji}=0$ & black line \\ \hline
		$A_{\mathcal{S}}=0$ &   ${}^{\mathcal{S}}C_{ji}=0$ & green line \\ \hline
		$A_{\mathcal{T}}=0$ &   ${}^{\mathcal{T}}C=0$ & blue line \\ \hline
		$A_{\mathcal{V}}=A_{\mathcal{A}}=0$ & ${}^{\mathcal{V}}C_{i}={}^{\mathcal{A}}C_{ji}=0$ & turquoise point \\ \hline
		$A_{\mathcal{A}}=A_{\mathcal{S}}=0$ &  ${}^{\mathcal{A}}C_{ji}={}^{\mathcal{S}}C_{ji}=0$ & purple point (center) \\ \hline
		$A_{\mathcal{A}}=A_{\mathcal{T}}=0$ &  ${}^{\mathcal{A}}C_{ji}={}^{\mathcal{T}}C=0$ & orange point \\ \hline
		$A_{\mathcal{V}}=A_{\mathcal{S}}=A_{\mathcal{T}}=0$ &  ${}^{\mathcal{V}}C_{i}={}^{\mathcal{S}}C_{ji}={}^{\mathcal{T}}C=0$ & gray points (perimeter)\footnotemark \\ \hline
	\end{tabular}
\end{center}
\footnotetext{This is actually only one point in the parameter space, since antipodal points on the perimeter correspond to the same theory.}

In order to understand the degrees of freedom and derive the full Hamiltonian of the theory, we would need to calculate the Poisson brackets and deduce whether they are first or second class constraints and if more constraints appear (secondary, tertiary etc). For teleparallel equivalence to general relativity this has already been done in \cite{Maluf:1994ji,Maluf:2000ag,Blagojevic:2000qs,Maluf:2001rg,Okolow:2011nq,Okolow:2013lwa,Ferraro:2016wht,Nester:2017wau} and it was found that the dynamical equivalent Hamiltonian to TEGR can be expressed with the help of two sets of Lagrange multipliers, ${}^\mathcal{V}\lambda^i$ and ${}^\mathcal{A}\lambda^{ij}$, as
\begin{align}\label{eq:Htegr}
\begin{split}
H_{TEGR} &= \sqrt{h} \Big({}^\mathcal{V}\lambda^i\ {}^\mathcal{V}C_i + {}^\mathcal{A}\lambda^{ij}\ {}^\mathcal{A}C_{ij}\Big) + D_i[\pi_A{}^i(\alpha \xi^{A}+\beta^{j}\theta^A{}_j )] \\
&- \alpha \sqrt{h} \Big( \frac{1}{4}{}^\mathcal{S}C_{ij}\ {}^\mathcal{S}C^{ij} - \frac{3}{8} {}^\mathcal{T}C\ {}^\mathcal{T}C +  {}^{3}\mathbb{T}  + \frac{\xi^A D_i\pi_A{}^i}{\sqrt{h}} \Big) - \beta^k\Big(T^A{}_{jk}\pi_A{}^j + \theta^A{}_k D_{i}\pi_A{}^i\Big)\,.
\end{split}
\end{align}

In the future we aim to derive the dynamically equivalent Hamiltonians for all nine classes we identified among the NGR theories of gravity. By introducing additional Lagrange multipliers ${}^\mathcal{S}\lambda^{ij}$ and ${}^\mathcal{T}\lambda^{ij}$ in the short-hand notation
\begin{align}
{}^\mathcal{V} H =
\begin{cases}
\alpha \sqrt{h}\ \frac{{}^{\mathcal{V}}C_i {}^{\mathcal{V}}C^i}{4 A_{\mathcal{V}}}& \textrm{ for } {}^{\mathcal{V}}A \neq 0\\
\sqrt{h}\ {}^\mathcal{V}\lambda_i {}^{\mathcal{V}}C^i & \textrm{ for } {}^{\mathcal{V}}A = 0\,,
\end{cases}\quad
{}^\mathcal{A} H =
\begin{cases}
-\alpha \sqrt{h}\ \frac{{}^{\mathcal{A}}C_{ij} {}^{\mathcal{A}}C^{ij}}{4 A_{\mathcal{A}}}& \textrm{ for } {}^{\mathcal{A}}A \neq 0\\
\sqrt{h}\ {}^\mathcal{A}\lambda_{ij} {}^{\mathcal{A}}C^{ij} & \textrm{ for } {}^{\mathcal{A}}A = 0\,,
\end{cases}\\
{}^\mathcal{S} H =
\begin{cases}
-\alpha \sqrt{h}\ \frac{{}^{\mathcal{S}}C_{ij} {}^{\mathcal{S}}C^{ij}}{4 A_{\mathcal{S}}}& \textrm{ for } {}^{\mathcal{S}}A \neq 0\\
\sqrt{h}\ {}^\mathcal{S}\lambda_{ij} {}^{\mathcal{S}}C^{ij} & \textrm{ for } {}^{\mathcal{S}}A = 0\,,
\end{cases}\quad
{}^\mathcal{T} H =
\begin{cases}
-\alpha \sqrt{h}\ \frac{3{}^{\mathcal{T}}C_{ij} {}^{\mathcal{T}}C^{ij}}{4 A_{\mathcal{T}}}& \textrm{ for } {}^{\mathcal{T}}A \neq 0\\
\sqrt{h}\ {}^\mathcal{T}\lambda_{ij} {}^{\mathcal{T}}C^{ij} & \textrm{ for } {}^{\mathcal{T}}A = 0\,,
\end{cases}
\end{align}
we can display a first step towards the dynamical Hamiltonians
\begin{align}\label{eq:Hdyn}
\begin{split}
H &=  \Big({}^\mathcal{V}H + {}^\mathcal{A}H + {}^\mathcal{S}H + {}^\mathcal{T}H \Big) - \alpha \Big(\sqrt{h}\ {}^{3}\mathbb{T}  - \xi^A D_i\pi_A{}^i \Big) - \beta^k\Big(T^A{}_{jk}\pi_A{}^j + \theta^A{}_k D_{i}\pi_A{}^i\Big) + D_i[\pi_A{}^i(\alpha \xi^{A}+\beta^{j}\theta^A{}_j )]\\
&+ \textrm{secondary-, tertiary-, \ldots{} constraints}\,.
\end{split}
\end{align}
However, the list of secondary-, tertiary-, \ldots{} constraints, which have to be added in addition, has to be investigated separately for the nine classes we derived. Even within a single class there may appear different constraint algebras. For example, in the class with all \(A_I\) being nonzero, the Poisson bracket of the Hamilton constraint with itself in general generates new constraints since the Poisson brackets of the Hamiltonian and momenta constraints do not form a closed algebra. However, for particular values of the parameters the terms which cause this behavior are absent from the action, thus allowing the Poisson brackets to close~\cite{Okolow:2011np}. Due to the lengthiness of the calculations even in seemingly simple cases such as TEGR~\cite{Okolow:2013lwa} we present these studies in separate articles. Another potential issue that must receive attention is the possible bifurcation of constraints, i.e., the situation where the closing or non-closing of the Poisson brackets depends on the particular values of the fields, as found in previous studies \cite{Chen:1998ad}, which we plan to investigate in detail in further work.

Before we conclude this article we like to add one more remark on the gauge fixing. The Hamiltonian we obtained is derived in the Weitzenb\"ock gauge. To remove the gauge fixing and to reintroduce the variables $\Lambda$ and $\hat \pi$, which we removed in the course of the discussion in section \ref{sec:momenta}, the following two steps have to be performed. First replace the Levi-Civita covariant derivatives $D_i$ in equation \eqref{eq:Hdyn} by a total covariant derivative $\mathfrak{D}_i$ which also acts on the Lorentz indices of the objects appearing,
\begin{align}
	D_i \pi_A{}^j \rightarrow \mathfrak{D}_i \pi_A{}^j = D_i \pi_A{}^j - \omega^B{}_{Ai}\pi_B{}^j\,,
\end{align}
and, second, add the constraint \eqref{conjugatemomentaauxillaryrel} with the help of a Lagrange multiplier. The result is a gauge invariant Hamiltonian depending on the field variables $\alpha,\ \beta^i, \theta^A{}_i, \pi_A{}^i$, and $\Lambda^A{}_B$ as well as $\hat \pi^{AB}$.

\section{Conclusion}\label{sec:conclusion}
We have derived a closed form for the kinematic Hamiltonian of \emph{new general relativity} theories of gravity, starting from its Lagrangian formulation including the teleparallel spin connection. The latter we implemented explicitly in terms of local Lorentz transformations, thus avoiding the need for Lagrange multipliers in the action. We found that the canonical momenta for the spin connection are not independent and can fully be expressed in terms of the momenta for the tetrad. Further, only the 12 spatial components of the tetrads have non-vanishing momenta, while the 4 temporal components can be expressed in terms of the ADM variables lapse and shift, whose momenta vanish identically. We have shown that it is not possible to invert the relation between the time derivatives of the spatial tetrad components and their conjugate momenta, which results in the appearance of up to four types of further primary constraints, depending on the choice of parameters defining the theory. We find that the family of NGR theories is divided into nine different classes, which are distinguished by the presence or absence of these primary constraints. We visualized the locations of these nine classes in the parameter space of the theory, and identified a prototype of a dynamically equivalent Hamiltonian for the different classes, which serves as a starting point for the continuation towards a complete systematic Hamiltonian analysis of NGR.

Our results invite further investigations in various directions. The most logical next step is the calculation of the Poisson brackets for all possible constraints. This will show under which circumstances the constraint algebra closes, and under which circumstances additional constraints must be included, and finally lead to the full, dynamical Hamiltonian. It should be noted that the calculation of the Poisson brackets is straightforward, although it can be very lengthy, even in the case of TEGR~\cite{Okolow:2013lwa}. Naively, the unconstrained case would be the easiest, since it involves the least number of constraints to calculate Poisson brackets with. However, the Poisson brackets do not form a closed algebra, hence are not first class, except for special cases~\cite{Okolow:2011np}, and thus generate further secondary constraints. Another class of new general relativity theories of particular interest besides general relativity is the one where only the vector constraint $A_{\mathcal{V}} = 0$ is imposed. It has been argued that this constraint is necessary in order to avoid the appearance of ghosts at the linearized level \cite{Kuhfuss1986,VanNieuwenhuizen:1973fi}. The constraint algebra has been worked out for this case, and it turns out that also in this case the constraints are not first class, so that secondary constraints appear \cite{Cheng:1988zg}.

An important result which we expect from the aforementioned further work on the constraint algebra is the number of degrees of freedom for general parameters of new general relativity. A hint towards the existence of further degrees of freedom compared to TEGR comes from comparing the degrees of freedom in new general relativity with the number of polarization modes of gravitational waves in the Newman-Penrose formalism~\cite{Hohmann:2018wxu}. This result gives a lower bound of the number of degrees of freedom, since the polarization modes which appear in the linearized theory must come from the fundamental degrees of freedom in the complete nonlinear theory. Once the full Hamiltonian is derived, it can be compared with the propagators presented in \cite{Koivisto:2018loq}. Results for a systematic categorization of theoretical pathologies (tachyons and ghosts) in a large class of theories including NGR was recently presented in \cite{Lin:2018awc}. Future work could consist of confirming their results using the Hamiltonian analysis and getting guidance in which theories are mostly motivated and perform the full-fledged Hamiltonian analysis in these cases.

The full dynamical Hamiltonian would also be useful for further tests of NGR with observations, in particular considering gravitational waves. The results we presented here show that the vicinity of TEGR in the parameter space, which is known to be compatible with post-Newtonian observations in the solar system~\cite{PhysRevD.19.3524}, is composed out of different classes of possible constraint algebras. Studying their Hamiltonian dynamics one may expect new results on the generation of gravitational waves in these theories, from which tighter bounds on the NGR parameters would be obtained.

\begin{acknowledgments}
The authors are grateful to Martin Kr\v{s}\v{s}\'ak for numerous discussions and pointing out references and to María José Guzmán for commenting on a previous version of this article. They were supported by the Estonian Research Council through the Institutional Research Funding project IUT02-27 and the Personal Research Funding project PUT790 (start-up project), as well as by the European Regional Development Fund through the Center of Excellence TK133 ``The Dark Side of the Universe''.
\end{acknowledgments}

\appendix

\section{Hamiltonian analysis without gauge fixing}
\label{Withoutgauge}
Looking at equation \eqref{conj} and noting that the conjugate momenta are related to each other via an algebraic equation \eqref{conjugatemomentaauxillaryrel} it at first seems like it is impossible to solve the velocities for momenta. However, there is a way to attack this problem and successfully derive the Hamiltonian. First, we note that equation \eqref{invertvelocity} before fixing the gauge becomes
\begin{align}
\label{invertvelocityungauged}
S_A{}^i= M^i{}_A{}^j{}_B \left(\dot{\theta}^B{}_j -\left(\Lambda^{-1}\right)^D{}_{C}\theta^C{}_j\dot{\Lambda}^B{}_D \right)
=M^i{}_A{}^j{}_B \Lambda^B{}_D \partial_0\left(\theta^C{}_j (\Lambda^{-1})^D{}_C \right)\,,
\end{align}
with
\begin{align}\label{eq:sourceungauged}
\begin{split}
S_A{}^i[\alpha,\beta,\theta^A{}_i,\pi_A{}^i]&=\frac{\alpha}{\sqrt{h}} \pi_A{}^i + \big[\Lambda^B{}_D D_{k}\left[ \left(\alpha \xi^{C} + \beta^{m}\theta^C{}_m \right)\left(\Lambda^{-1}\right)^D{}_C \right] - T^B{}_{kl}\beta^{l} \big] M^i{}_A{}^k{}_B
\\
&- 2 \alpha T^B{}_{kl} h^{ik}( c_{2}\xi_{B}\theta_A{}^l+c_{3}\xi_{A}\theta_B{}^l).
\end{split}
\end{align}
In the Lagrangian, velocities only appear from terms of the structure
\begin{align}
	T^B{}_{0j}=\Lambda^B{}_D\partial_0\left(\theta^C{}_j\left(\Lambda^{-1}\right)^D{}_C\right)-\Lambda^B{}_DD_j\left[\left(\alpha\xi^C+\beta^m\theta^C{}_m\right)\left(\Lambda^{-1}\right)^D{}_C\right].
\end{align}
Hence, the velocities in the Lagrangian appear exactly as in equation \eqref{invertvelocityungauged}. This means that we can get rid of all velocities and express them in terms of conjugate momenta by applying $\left(M^{-1}\right)^{A \ C}_{\ i \ k}$ on both sides of equation \eqref{invertvelocityungauged}, where we have used the same decomposition of the Weitzenböck tetrad $\dot{\tilde{\theta}}^A{}_i=\partial_0\left(\theta^B{}_i \left(\Lambda^{-1}\right)^A{}_B\right)$ as in equation \eqref{veldec} into irreducible parts.
\\ \indent Second, we need to write down the Hamiltonian together with its primary constraints. The algebraic relation between the conjugate momenta is a primary constraint and needs to be added. The Hamiltonian is then by definition
\begin{align}
	\label{ungaugedHamiltonian}
	H=\pi_A{}^i\dot{\theta}^A{}_i+\hat{\pi}^{AB}a_{AB}-L(\theta^A{}_i,\pi_A{}^i)-{}^\pi\lambda^{A}{}_{B}\left(\hat \pi^B{}_A +\pi_A{}^i\eta^{B[N}\theta^{M]}{}_i\right),
\end{align}
which is the gauge independent correspondence to equation \eqref{eq:HNGR}. Using the equation imposed by the Lagrange multiplier to express all conjugate momenta solely in the conjugate momenta with respect to the spatial tetrad field $\pi_A{}^i$ we get that the Hamiltonian is of the form
\begin{align}
\label{}
H=\pi_A{}^i\Lambda^A{}_B\partial_0\left(\theta^C{}_i\left(\Lambda^{-1}\right)^B{}_C\right)-L\left[\alpha,\beta,\theta^A{}_i,\pi_A{}^i,\Lambda^A{}_B\right]-{}^\pi\lambda^{A}{}_{B}\left(\hat \pi^B{}_A +\pi_A{}^i\eta^{B[N}\theta^{M]}{}_i\right).
\end{align}
From this we can see that the Hamiltonian can be expressed in canonical variables without gauge fixing. By using equation \eqref{invertvelocityungauged} we get
\begin{align}
\begin{split}
H[\alpha,\beta,\theta^A{}_i,\pi_A{}^i,\Lambda^A{}_B,\hat{\pi}^B{}_A]&=\pi_A{}^i\left(M^{-1}\right)^{A \ C}_{\ i \ k}S_C{}^k[\alpha,\beta,\theta^A{}_i,\pi_A{}^i]-L\left[\alpha,\beta,\theta^A{}_i,\pi_A{}^i,\Lambda^A{}_B\right]\\
&-{}^\pi\lambda^{A}{}_{B}\left(\hat \pi^B{}_A +\pi_A{}^i\eta^{B[N}\theta^{M]}{}_i\right).
\end{split}
\end{align}

\bibliographystyle{utphys}
\bibliography{NGRADMBlixt2}
\end{document}